\documentclass[%
 reprint,
superscriptaddress,
bibnotes,
 amsmath,amssymb,
 aps,
prb,
]{revtex4-2}

\usepackage{graphicx}
\usepackage{dcolumn}
\usepackage{bm}
\usepackage{caption}
\usepackage{float}
\usepackage{amsmath}
\usepackage{animate}
\usepackage{hyperref}
\usepackage{chemformula}
\usepackage{bbm}

\pdfoutput=1
\DeclareCaptionFormat{myformat}{\hspace{1.5em}#1#2#3\par}
\begin{document}
 

\title{1/3 and other magnetization plateaus in a quasi-one-dimensional Ising magnet $\mathbf{TbTi_3Bi_4}$ with zigzag spin chain}

\author{Kaizhen Guo}
\affiliation{
International Center for Quantum Materials, School of Physics, Peking University, Beijing 100871, China
}%

\author{Zeyu Ma}
\affiliation{
	International Center for Quantum Materials, School of Physics, Peking University, Beijing 100871, China
}%

\author{Hongxiong Liu}
\affiliation{
	Beijing National Laboratory for Condensed Matter Physics and Institute of Physics, Chinese Academy of Sciences, Beijing 100190, China
}%
\affiliation{
	University of Chinese Academy of Sciences, Beijing 100049, China
}%

\author{Ziyang Wu}
\affiliation{
	Wuhan National High Magnetic Field Center and School of Physics, Huazhong University of Science and Technology, Wuhan 430074, China
}%

\author{Junfeng Wang}
\affiliation{
	Wuhan National High Magnetic Field Center and School of Physics, Huazhong University of Science and Technology, Wuhan 430074, China
}%

\author{Youguo Shi}
\affiliation{
	Beijing National Laboratory for Condensed Matter Physics and Institute of Physics, Chinese Academy of Sciences, Beijing 100190, China
}%
\affiliation{
	University of Chinese Academy of Sciences, Beijing 100049, China
}%

\author{Yuan Li}
\affiliation{
	International Center for Quantum Materials, School of Physics, Peking University, Beijing 100871, China
}%

\author{Shuang Jia}
\homepage{gwljiashuang@pku.edu.cn}
\affiliation{
International Center for Quantum Materials, School of Physics, Peking University, Beijing 100871, China
}%
\affiliation{
 Interdisciplinary Institute of Light-Element Quantum Materials and Research Center for Light-Element Advanced Materials, Peking University, Beijing 100871, China
}%

\date{\today}

\begin{abstract}

We report the magnetic properties of newly synthesized, single crystals of $\mathrm{TbTi_3Bi_4}$ whose crystal structure is highlighted by the stacking of terbium-based zigzag chains and titanium-based kagome lattices. 
This compound demonstrates extreme easy-axis magnetic anisotropy due to the crystalline-electric-field effect which aligns the $\mathrm{Tb^{3+}}$ moments along the zigzag chain direction.
As the result of the strong single-ion anisotropy and multiple magnetic interactions, $\mathrm{TbTi_3Bi_4}$ behaves as a quasi-one-dimensional Ising magnet with a remarkable antiferromagnetic ordering at $T_\mathrm{N} = 20.4$ K.
When a magnetic field is applied along the direction of the zigzag chain, multiple meta-magnetic transitions occur between 1/3 and other magnetization plateaus.
We have created a field-temperature phase diagram and mapped out the complex magnetic structures resulting from frustration.

\end{abstract}

\maketitle

\section{INTRODUCTION}

One-dimensional and quasi-one-dimensional quantum magnets have received significant attention because they can showcase various emerging collective behaviors and strong quantum fluctuations \cite{reviewSteiner,RevModPhys.93.025003}.
The presence of plateaus in the magnetization curves of low-dimensional spin systems, which generally occurs at rational fractions of the saturation magnetization, is one interesting characteristic \cite{AHonecker1999,PhysRevB.75.064413,PhysRevLett.101.177201,Shi2022,Okunishi,AASLAND1997187,NANDI2016121,PhysRevB.68.224422}.
The mechanisms of the fractional magnetization plateaus are varied, including dimerization, frustration and single-ion anisotropy \cite{AHonecker1999,PhysRevB.75.064413,PhysRevB.54.9862}.
The spin zigzag chain is one of the most fundamental models that includes the frustrating interaction in one dimension.
With only two parameters of magnetic interactions, the spin zigzag chain captures a variety of interesting behaviors induced by frustration.
The zigzag chain can exhibit an exotic phase diagram in an external magnetic field, including the magnetization plateau at $1/3$ of the full moment and the fractional excitations around the plateaus \cite{Okunishi,PhysRevB.68.224422}.
Understanding the phase diagram of the real materials with spin zigzag chains will help to understand the role of the magnetic interaction and frustration in quantum magnets.

The intermetallic $R\mathrm{Ti_3Bi_4}$ ($R$ = La-Nd, Sm-Gd, and Yb) are newly discovered isostructural family of $R\mathrm{V_3Sb_4}$ ($R$ = Eu and Yb) compounds \cite{PhysRevMaterials.7.064201,MOTOYAMA2018142,doi:10.1021/acs.chemmater.3c02289,chen2023tunable,jiang2023direct,guo2023magnetic,zheng2024anisotropic} that crystallize in an orthorhombic space group $Fmmm$ (No. 69).
Their crystal structure is highlighted by the stacking of rare-earth-based zigzag chains and titanium-based kagome lattices, reminiscent of the vanadium-based kagome lattice in $A\mathrm{V_3Sb_5}$ ($A$ = K, Rb, Cs) and $R\mathrm{V_6Sn_6}$, which play a crucial role in the unique intertwining charge density wave (CDW) order and superconducting ground state \cite{PhysRevMaterials.3.094407,PhysRevLett.125.247002,PhysRevMaterials.5.034801,Yin2021,PhysRevLett.127.266401,PhysRevLett.129.216402,PhysRevMaterials.6.105001,PhysRevMaterials.6.083401,PhysRevB.107.205151}. 
The physical properties of $R\mathrm{Ti_3Bi_4}$ are governed by the $R$ magnetism, ranging from nonmagnetic $\mathrm{(La,Yb)Ti_3Bi_4}$, possibly short-range ordering in $\mathrm{PrTi_3Bi_4}$, antiferromagnetic (AFM) ordering of the Heisenberg moments in $\mathrm{(Ce,Gd)Ti_3Bi_4}$, and ferromagnetic (FM) ordering in $\mathrm{(Nd,Sm,Eu)Ti_3Bi_4}$.
Current studies on rare earth compounds mainly focus on light $R$ and $\mathrm{GdTi_3Bi_4}$, while the existence of the heavier $R\mathrm{Ti_3Bi_4}$ is unknown, probably due to the structural instability caused by lanthanide contraction.
In the current study, we show the existence of isostructural $\mathrm{TbTi_3Bi_4}$ which  manifests the magnetic properties of a quasi-one-dimensional, Ising-like quantum magnet.

We have successfully obtained the high-quality single crystals of $\mathrm{TbTi_3Bi_4}$ from Bi flux.
Our magnetization and electric transport measurements on the single crystals of $\mathrm{TbTi_3Bi_4}$ unveiled an AFM ordering at $T_\mathrm{N} = 20.4$ K and strong magnetic anisotropy with the easy axis of the zigzag chain direction ($a$-axis).
By measuring the single crystals of $\mathrm{La_{0.972}Tb_{0.028}Ti_3Bi_4}$ we found that the magnetic moments of the $\mathrm{Tb^{3+}}$ ions are aligned along the $a$-axis direction due to the crystalline-electric-field (CEF) effect.
As a result of the single-ion anisotropy and multiple magnetic interactions, $\mathrm{TbTi_3Bi_4}$ behaves as a quasi-one-dimensional Ising magnet in which the Ising moments are aligned along the $a$-axis.
We observed multiple meta-magnetic transitions (MMTs) and 1/3 plateau on the magnetic isothermals at temperatures below $T_\mathrm{N}$ when a magnetic field is applied along the $a$ direction.  
The rich phase diagram, which includes multiple fractional magnetization plateaus and meta-stable cusps, suggests that $\mathrm{TbTi_3Bi_4}$ is a potential model quantum magnet for studying the diverse physics of the spin zigzag chain in the Ising limit.

\section{EXPERIMENTAL METHOD}

Single crystals of $\mathrm{TbTi_3Bi_4}$, $\mathrm{LaTi_3Bi_4}$, and $\mathrm{La_{0.972}Tb_{0.028}Ti_3Bi_4}$ were synthesized via the self-flux method. The starting elements of Tb (grains, 99.9\%), La (grains, 99.9\%), Ti (grains, 99.9\%), and Bi (shots, 99.9\%) with the molar ratio Tb: La: Ti: Bi = 1-$x$: $x$: 1: 20 ($x = 0, 0.95, 1$) were placed in an alumina Canfield crucible set (CCS) which is effective at preventing samples from contacting the silica wool \cite{doi:10.1080/14786435.2015.1122248}, and then sealed in a vacuum silica ampoule. The flux mixture was heated to $\mathrm{1000^\circ C}$, allowed to dwell for 12 h, then continuously cooled at a cooling rate of $\mathrm{5^\circ C/h}$. We annealed the mixture at $\mathrm{500^\circ C}$ for about 5 days and then finally centrifuged it to remove excess flux. 
It should be emphasized that long-time annealing at $\mathrm{500^\circ C}$ is necessary to obtain large single crystals of $\mathrm{TbTi_3Bi_4}$.
The yielded crystals are thick hexagonal plates with dimensions of $\mathrm {2\times 2\times 0.5~mm^3}$ (see Fig.~\ref{fig1} d). 
The crystal structure was confirmed by performing X-ray diffraction (XRD) measurements.
We collected the diffraction data for a single-crystalline plate at room temperature using a Rigaku Mini-flex 600 instrument, and the pattern displayed a preferential orientation of $(00l)$ ($l$ = even integer) reflections (Fig.~\ref{fig1} e), the same as other $R\mathrm{Ti_3Bi_4}$ compounds.
Because it is difficult to grind the crystals into powder, we performed single-crystal X-ray crystallographic analysis by using Bruker D8 Venture. The frames were integrated with the Bruker SAINT software package using a narrow-frame algorithm. The structure was solved and refined using the Bruker SHELXTL Software Package, using the space group $Fmmm$.
The single-crystal XRD data collection and refinement parameters for $\mathrm{TbTi_3Bi_4}$ are gathered in Tables~\ref{table1}, \ref{table2}, and \ref{table3}.

\begin{table}[h]
	\caption{\label{table1}Data collection and refinement parameters for $\mathrm{TbTi_3Bi_4}$ from single crystal X-ray diffraction.}
	\begin{tabular}{ll}
		\hline \hline
		Chemical formula, Z & $\mathrm{TbTi_3Bi_4}$, 8  \\
		Radiation, $\lambda(\mathrm{\AA})$ & Mo $\mathrm{K} \alpha(\lambda=0.71073)$  \\
		Temperature $(\mathrm{K})$ & $273(2)$  \\
		Formula weight (g/mol) & 1138.54 \\
		Crystal system & Orthorhombic \\
		Space group & $Fmmm$ \\
		Unit-cell dimensions & \\
		$a(\mathrm{\AA})$ & $5.8681(8)$ \\
		$b(\mathrm{\AA})$ & $10.3473(15)$ \\
		$c(\mathrm{\AA})$ & $24.785(4)$ \\
		$\alpha=\beta=\gamma\left({ }^{\circ}\right)$ & 90 \\
		Volume $\left(\mathrm{\AA^3}\right)$ & $1504.9(4)$ \\
		Absorption coefficient & 19.183 \\
		$F(000)$ & 3704 \\
		Density $\left(\mathrm{g} / \mathrm{cm}^3\right)$ & 10.050 \\
		Crystal size $\left(\mathrm{mm}^3\right)$ & $0.009 \times 0.034 \times 0.054$ \\
		$2 \theta$ range for data collection $\left({ }^{\circ}\right)$ & 1.64 to 30.49 \\
		Index ranges & $-8 \leqslant h \leqslant 7,-14 \leqslant k \leqslant 14$, \\
		& $-35 \leqslant l \leqslant 35$ \\
		Reflections collected & 5738 \\
		Independent reflections & $678\left[R_{\text {int }}=0.0744\right.$ \\
		& $\left.R_{\text {sigma }}=0.0383\right]$ \\
		Data/Retraints/Patameters & $678 / 0 / 29$ \\
		Goodness-of-fit on $F^2$ & 1.145 \\
		Final $R$ indexes (566 data)$[I \geqslant 2 \sigma(I)]$ & $R_{\mathrm{obs}}=0.0706$ \\
		& $\mathrm{w} R_{\mathrm{obs}}=0.2295$ \\
		Final $R$ indexes (all data) & $R_{\mathrm{all}}=0.0816$ \\
		& $\mathrm{w} R_{\mathrm{all}}=0.2380$ \\
		Largest diff. peak/hole $\left(e \mathrm{\AA^3}\right)$ & $8.112 /-5.903$ \\
		\hline \hline
		
	\end{tabular}
\end{table}

\begin{table}[h]
	\caption{\label{table2}Atomic coordinates and equivalent isotropic displacement parameters for $\mathrm{TbTi_3Bi_4}$.}
	\begin{tabular}{cccccc}
		\hline \hline Atom & $x / a$ & $y / b$ & $z / c$ & $O c c$. & $U(\mathrm{eq})\left(\AA^2\right)$ \\
		\hline $\mathrm{Tb}$ & 0 & 0.500000 & 0.69598(13) & 1.0 & $0.0196(7)$ \\
		Ti1 & 0.250000 & 0.250000 & 0.5962(3) & 1.0 & 0.0175(14) \\
		Ti2 & 0.500000 & 0.500000 & 0.5932(5) & 1.0 & 0.019(2) \\
		Bi1 & 0.500000 & 0.34098(14) & 0.69017(7) & 1.0 & 0.0180(5) \\
		Bi2 & 0.500000 & 0.33007(19) & 0.500000 & 1.0 & 0.0186(6) \\
		Bi3 & 0 & 0.500000 & 0.56978(10) & 1.0 & 0.0187(6) \\
		\hline \hline
	\end{tabular}
	 
\end{table}

\begin{table}[h]
	\caption{\label{table3}Anisotropic displacement parameters ($\mathrm{\AA^2 \times 10^{-2}}$) for $\mathrm{TbTi_3Bi_4}$ at room temperature from single-crystal XRD. The anisotropic atomic displacement factor exponent takes the form: $-2 \pi^2\left(h^2 a^2 U_{11}+k^2 b^2 U_{22}+\ldots+2 K l b c U_{23}\right)$.}
	\begin{tabular}{ccccccc}
		\hline \hline Atom & $U_{11} $ & $U_{22} $ & $U_{33} $ & $U_{23} $ & $U_{13} $ & $U_{12} $ \\
		\hline Tb & $1.37(12)$ & $1.36(12)$ & $3.14(17)$ & 0 & 0 & 0 \\
		Ti1 & $1.20(3)$ & $1.10(3)$ & $3.0(4)$ & 0 & 0 & -0.1(2) \\
		Ti2 & $1.30(4)$ & $1.80(5)$ & $2.5(6)$ & 0 & 0 & 0 \\
		Bi1 & $1.30(8)$ & $1.10(8)$ & $2.99(10)$ & -0.02(5) & 0 & 0 \\
		Bi2 & $1.47(10)$ & $1.21(10)$ & $2.90(12)$ & 0 & 0 & 0 \\
		Bi3 & $1.26(9)$ & $1.07(9)$ & $3.29(13)$ & 0 & 0 & 0 \\
		\hline
	\end{tabular}

\end{table}

Resistance and heat capacity measurements were carried out using Quantum Design Physical Properties Measurement System (PPMS). The resistance measurement was performed using a standard four-probe method with the current flowing perpendicular to the $c$-axis.
Magnetization measurements were carried out using a Quantum Design Magnetic Properties Measurement Systems (MPMS-3). The isothermal magnetization up to 500 kOe was measured in a pulsed magnetic field at Wuhan National High Magnetic Field Center.

\section{RESULTS}

\subsection{Crystal Structure}

\begin{figure}[!htbp]
	\centering
	\includegraphics[width=\linewidth]{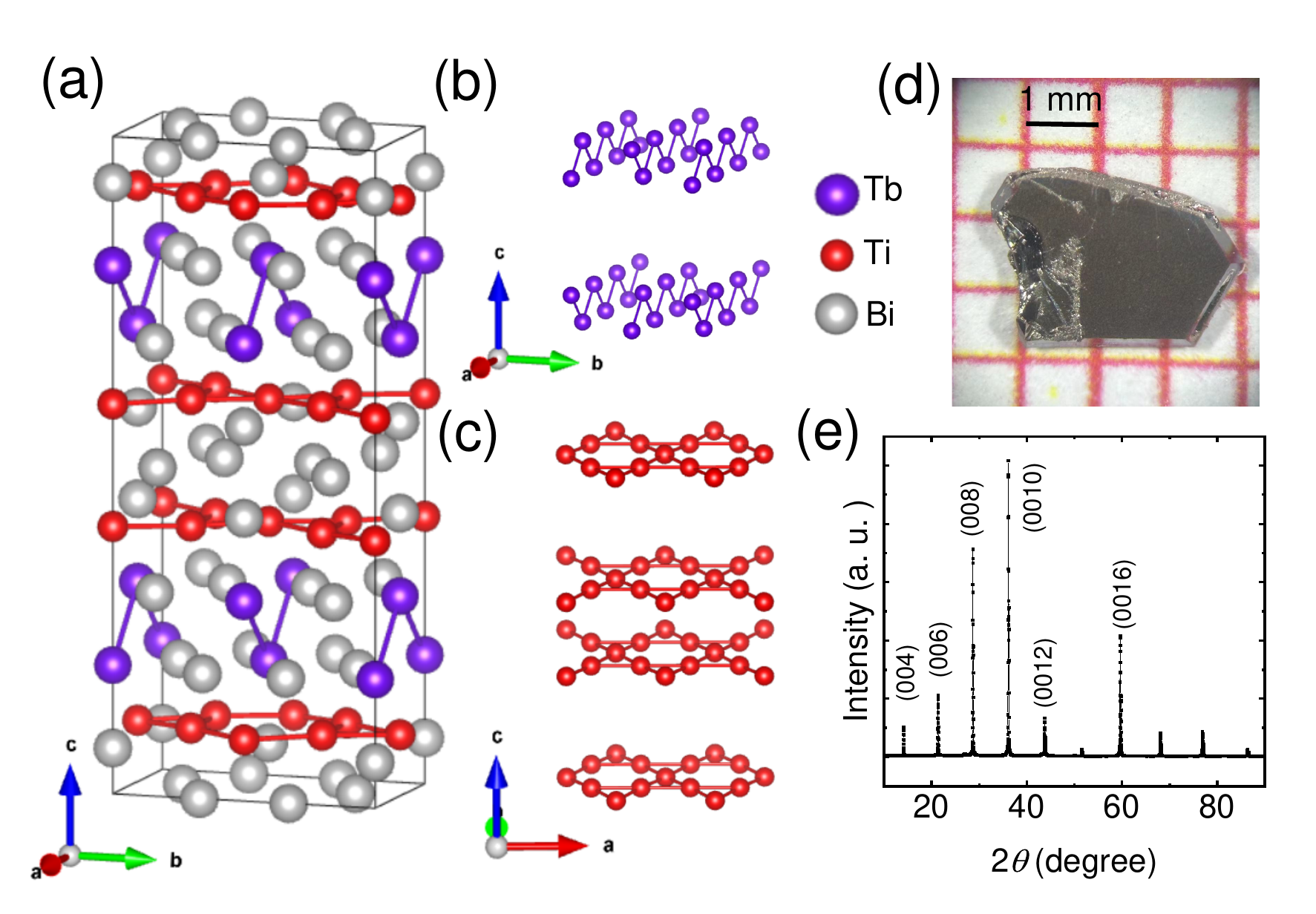}
	\caption{\label{fig1} Crystal structure of $\mathrm{TbTi_3Bi_4}$. (a) A unit cell of $\mathrm{TbTi_3Bi_4}$ where the purple, red, and grey colored solid spheres denote the Tb, Ti, and Bi atoms, respectively. (b) The sublattice of Tb atoms, where the purple line segments emphasize the nearest neighbor bonds in the zigzag chain. (c) The bilayer, Ti-based kagome lattice, which is slightly distorted and not co-planar. (d) A photograph of a single crystal of $\mathrm{TbTi_3Bi_4}$. (e) XRD pattern of as-grown $\mathrm{TbTi_3Bi_4}$ single crystals, showing $(00l)$ ($l$ = even integers) reflections. }
\end{figure}

$\mathrm{TbTi_3Bi_4}$ crystallizes in a layered, orthorhombic structure which is the same as that of other $R\mathrm{Ti_3Bi_4}$ compounds (space group $Fmmm$ No. 69).  
Detailed crystal structure information is listed in Table \ref{table1}.
The structure of $\mathrm{TbTi_3Bi_4}$ is highlighted by bilayer, distorted Ti-based kagome lattice, and bilayer Tb-based triangle lattice (Fig.~\ref{fig1} b and c).
The double layers of the Tb atoms are shifted by $a/2$ along the crystallographic $a$ direction, leading to Tb-based zigzag chains with the nearest Tb-Tb distance of 3.97 $\mathrm{\AA}$  (Fig.~\ref{fig1} b).
On the other hand, the inter-chain distance of 5.17 $\mathrm{\AA}$ along the $b$ direction and 9.72 $\mathrm{\AA}$ along the $c$ direction are much larger, indicating a dominated quasi-one-dimensional magnetic interaction.

\subsection{Antiferromagnetic Order}

\begin{figure}[!htbp]
	\centering
	\includegraphics[width=\linewidth]{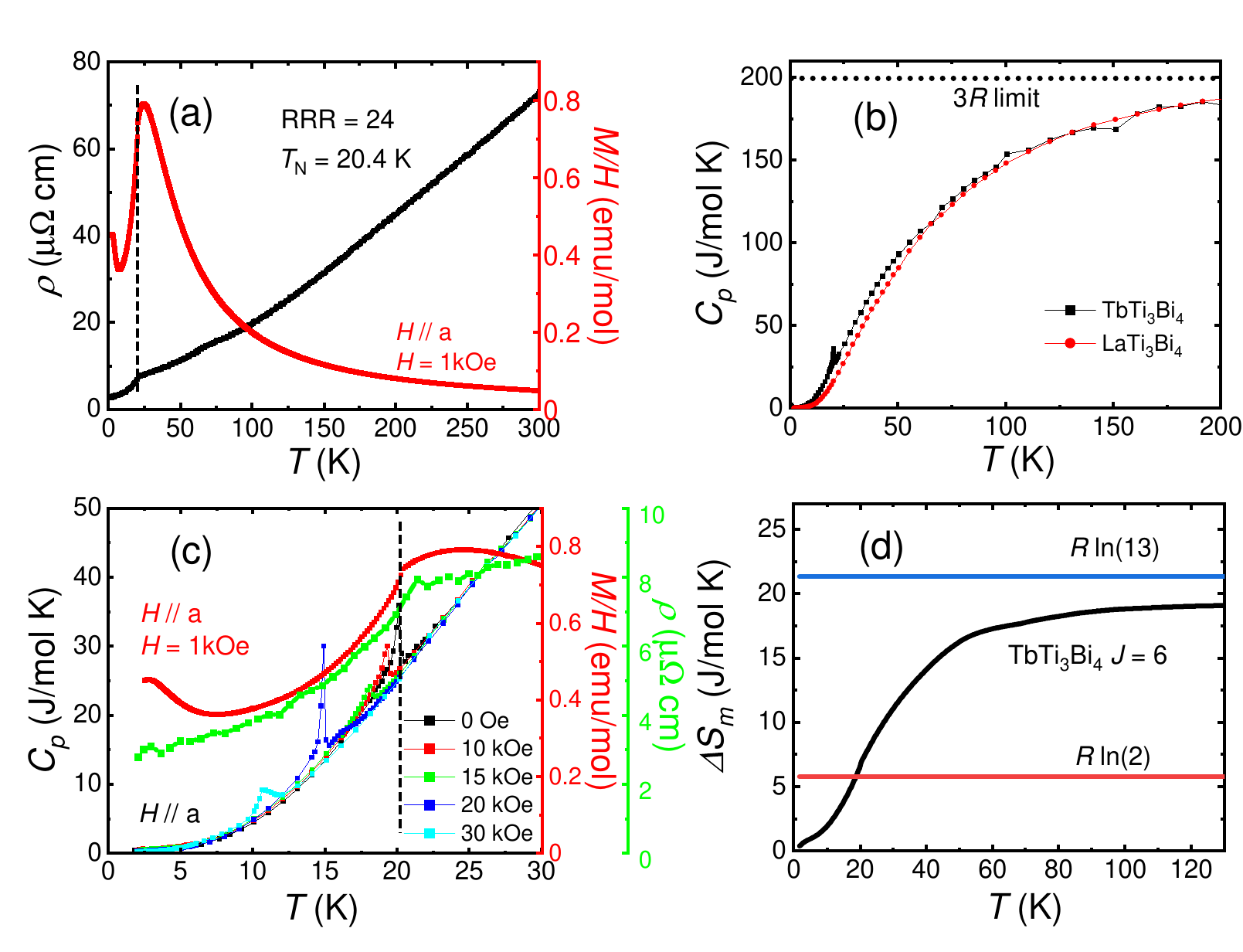}
	\caption{\label{fig2} Antiferromagnetism of $\mathrm{TbTi_3Bi_4}$. (a) Left axis: Temperature-dependent resistivity of $\mathrm{TbTi_3Bi_4}$ in zero field. Right axis: The magnetic susceptibility in a magnetic field of $H$ = 1~kOe for $H \parallel a$. The vertical dot line indicates a clear phase-transition appears at 20.4~K. (b) Specific heat $C_p$ versus temperature. The black curve is $\mathrm{TbTi_3Bi_4}$ and the red curve is $\mathrm{LaTi_3Bi_4}$ for comparison. The horizontal dot line represents the $3R$ limit according to the Debye model. (c) Left axis: Specific heat $C_p$ versus temperature in different magnetic fields along the $a$-axis. Right axis: The magnetic susceptibility in a magnetic field of $H$ = 1~kOe for $H \parallel a$. (d) Temperature-dependent magnetic entropy. Here we subtract the phonon contribution by subtracting the specific heat of $\mathrm{LaTi_3Bi_4}$.}
\end{figure}

We present the temperature-dependent magnetic susceptibility $\chi (T)$ for a single crystal of $\mathrm{TbTi_3Bi_4}$ in a magnetic field of 1 kOe along the $a$-axis in Fig.~\ref{fig2} a.
The curve of $\chi (T)$ displays a broad maximum at approximately 25 K and a slight slope change at 20.4 K, followed by a weak upturn below 5 K with decreasing temperature.
The temperature dependence of the resistivity ($\rho (T)$) shows a metallic profile in zero field with a significant slope change at 20.4 K (Fig.~\ref{fig2} a, black curve), indicating a magnetic ordering onsite.
Figure \ref{fig2} b shows the specific heat ($C_p$) for $\mathrm{TbTi_3Bi_4}$ and $\mathrm{LaTi_3Bi_4}$ from 1.8 to 200 K in zero magnetic field. 
As a nonmagnetic counterpart, the $C_p$ of $\mathrm{LaTi_3Bi_4}$ exhibits no signal of phase transition, similar to what was reported in Refs.~\cite{MOTOYAMA2018142,doi:10.1021/acs.chemmater.3c02289}.
On the other hand, the $C_p$ curve of $\mathrm{TbTi_3Bi_4}$ is similar to that of $\mathrm{LaTi_3Bi_4}$, except for a sharp and narrow $\lambda$-shaped peak at approximately 20~K.
At high temperatures, both curves tend to be saturated and approach the $3R$ limit.

We zoom in the temperature-dependent specific heat, resistivity and magnetic susceptibility on the same temperature scale below 30 K, as seen in Fig.~\ref{fig2} c.
We use the vertical dotted line to indicate the position of the AFM magnetic ordering temperature ($T_{\mathrm{N}}$) to show the consistency among the $\chi(T)$,  $C_p$, and $\rho (T)$.
To confirm the nature of the AFM ordering, we measured the $C_p$ under different magnetic fields along the $a$-axis.
The $\lambda$-shaped peak in specific heat initially becomes broader and moves to the lower temperature when the magnetic fields are 10 and 15~kOe.
However, when the magnetic field is 20 kOe, the peak becomes narrow again at 15~K, which indicates a different magnetic structure in this field.
Under higher magnetic field, the peak continuously moves towards the lower temperature and becomes broader again.

We inferred the magnetic entropy ($S_m$) by integrating $C_p/T$ after subtracting $C_p/T$ of $\mathrm{LaTi_3Bi_4}$, which was considered as phonon contribution (Fig.~\ref{fig2} d). It reaches  $R \ln2$ at $T_\mathrm{N} = 20.4$~K and tends to saturate above 50 K, which is close to $R\ln13 $, the magnetic entropy for the Hund’s rule ground state of $\mathrm{Tb^{3+}}$. This observation reveals that the local moment associated with the Hund’s rule ground state of $\mathrm{Tb^{3+}}$ has undergone an energy splitting at temperatures higher than $T_{\mathrm{N}}$ while the AFM ordering is associated with a doublet, although $\mathrm{Tb^{3+}}$ is a non-Kramers ion.

\begin{figure}[!htbp]
	\centering
	\includegraphics[width=\linewidth]{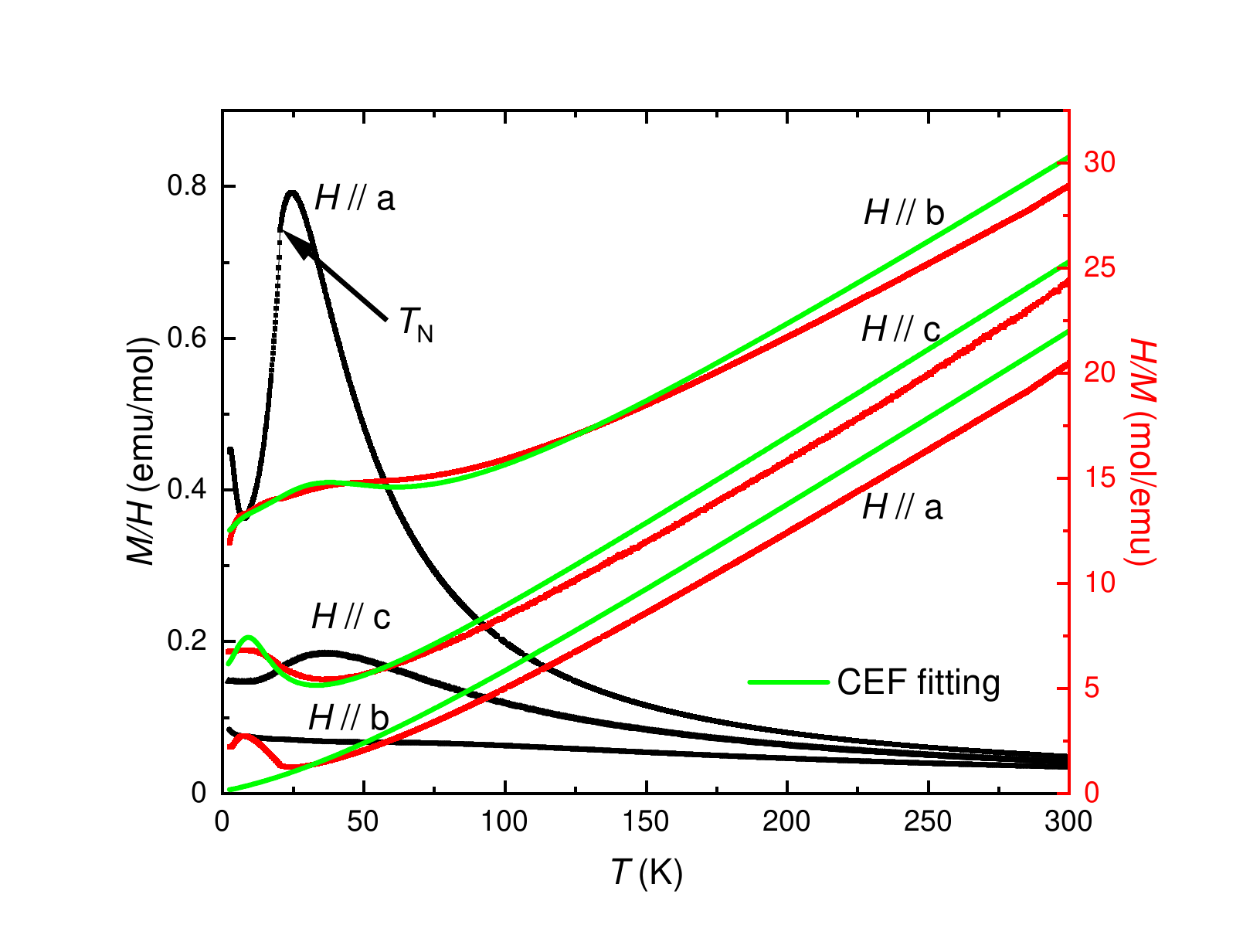}
	\caption{\label{fig3} The magnetic susceptibility (black curves) and its inverse (red curves) in a magnetic field of $H$ = 1 kOe for $H \parallel a$-, $b$-, and $c$-axes, respectively. The green curves represent the result of CEF fitting. }
\end{figure}

Figure~\ref{fig3} demonstrates anisotropic temperature-dependent magnetic susceptibility $\chi (T)$ when a field of $H$ = 1 kOe was applied along the directions of $a$-, $b$-, and $c$-axes, respectively.
The easy axis is the $a$-axis while the hard axis is the in-plane $b$-axis, which reveals one-dimensional instead of two-dimensional magnetic anisotropy.
While the $\chi (T)$ curve has no significant feature when $H\parallel b$, it has a broad peak at around 35~K when $H\parallel c$, which is distinctive from the broad peak at around 25~K when $H\parallel a$.  
When the temperatures are higher than 200~K, the inverse susceptibility curves ($1/\chi (T)$) are nearly straight and parallel to each other.
We fitted the high-temperature region by using the Curie-Weiss law, $\chi(T) = C/(T-\theta_{\mathrm{P}})$, where  $C=N_{\mathrm{A}}\mu_{eff}^2{\mu_{\mathrm{B}}^2}/3k_{\mathrm{B}}$ and $\theta_{\mathrm{P}}$ is the Weiss temperature. Here $N_{\mathrm{A}}$ is the Avogadro number, $\mu_{\mathrm{B}}$ is the Bohr magneton, and $k_{\mathrm{B}}$ is the Boltzmann constant.
The results are $\mu_{eff}^{a} = 9.99~\mu_{\mathrm{B}}$, $\theta_{\mathrm{P}}^{a} = 45.3~\mathrm{K}$ for $H\parallel a$-axis, and $\mu_{eff}^{b} = 10.55~\mu_{\mathrm{B}}$, $\theta_{\mathrm{P}}^{b} = -101.6$~K for $H\parallel b$-axis, and $\mu_{eff}^{c} = 9.68~\mu_{\mathrm{B}}$, $\theta_{\mathrm{P}}^{c} = 15.6$~K for $H\parallel c$-axis.
The obtained effective moments are close to the value for the Hund's rule ground state of $\mathrm{Tb^{3+}}$, $\mu_{eff} = 9.72~\mu_{\mathrm{B}}$ and the average Weiss temperature is $\theta_{\mathrm{P}} = (\theta_{\mathrm{P}}^{a} + \theta_{\mathrm{P}}^{b} + \theta_{\mathrm{P}}^{c})/3=-13.6~\mathrm{K}$, implying an AFM interaction of $\mathrm{Tb^{3+}}$ in average at high temperatures.
The large difference between the positive Weiss temperature $\theta_{\mathrm{P}}^{a} = 45.3~\mathrm{K}$ and the $\mathrm{N\acute{e}el}$ temperature $T_{\mathrm{N}}= 20.4 ~\mathrm{K}$ indicates that a strong magnetic frustration may play a role in the magnetic ordering.

The $\chi (T)$ curves of $\mathrm{TbTi_3Bi_4}$ display anisotropy and temperature dependence which can be explained by a dominant CEF effect with relatively weak magnetic interaction.
Assuming a single-ion Hamiltonian with an orthorhombic coordination CEF, we calculated $1/\chi (T)$ as the green curves in Fig.~\ref{fig3}.
The calculated results are consistent with the experimental data for $H \parallel b$ and $c$ in the entire temperature range but deviate from the data for $H \parallel a$ when the temperatures are below 30~K.
It is worth noting that the calculation follows the broad peak at about 35~K for $H \parallel c$ but completely deviates from the broad peak at 25~K for $H \parallel a$, which is due to a magnetic interaction and is characteristic of low-dimensional spin systems  \cite{crossover}. 
All the above data indicate that  $\mathrm{TbTi_3Bi_4}$ is a quasi-one-dimensional magnetic system in which the CEF effect plays an important role.

\subsection{Anisotropic magnetization in $\mathbf{La_{0.972}Tb_{0.028}Ti_3Bi_4}$}

\begin{figure}[!htbp]
	\centering
	\includegraphics[width=\linewidth]{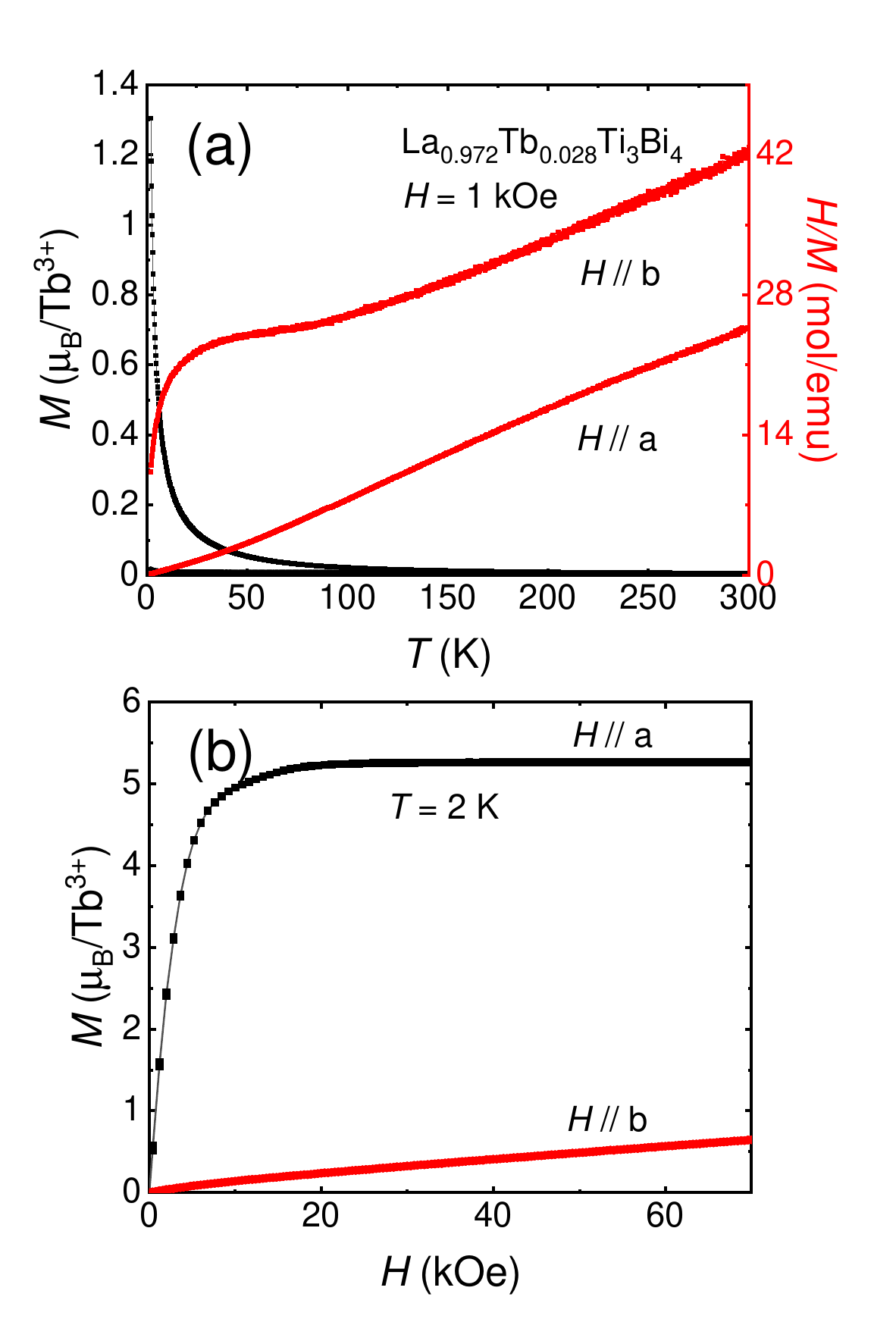}
	\caption{\label{fig4} The magnetic properties of $\mathrm{La_{0.972}Tb_{0.028}Ti_3Bi_4}$. (a) Left axis: Temperature-dependent magnetization in a magnetic field of $H$ = 1~kOe for $H \parallel a$- and $b$-axes. Right axis: The inverse magnetic susceptibility in a magnetic field of $H$ = 1~kOe for $H \parallel a$- and $b$-axes. (b) Field-dependent magnetization at 2~K.}
\end{figure}

We conducted a study to better understand the quasi-one-dimensional magnetic anisotropy in $\mathrm{TbTi_3Bi_4}$ by growing a single-crystalline alloy of $\mathrm{La}_{1-x}\mathrm{Tb}_{x}\mathrm{Ti}_3\mathrm{Bi}_4$.
We replaced most of the magnetic $\mathrm{Tb^{3+}}$ ions with non-magnetic $\mathrm{La^{3+}}$ ions, which results in a negligible magnetic interaction between the Tb moments.
As shown in Fig.~\ref{fig4} a, the magnetic susceptibility strictly followed the CEF fitting curves for $H \parallel a$ and $b$, with no evidence of magnetic ordering.
By considering the effective moments for $H \parallel a$ and $b$ fitted by the Curie-Weiss law, we inferred the concentration $x$ to be $0.028$ in the crystal.
The Weiss temperatures are 6.2~K for $H \parallel a$ and -192~K for  $H \parallel b$, indicating a weak magnetic interaction and a strong CEF effect.
We observed that the $\chi (T)$ for $H \parallel b$-axis behaves nearly the same as that of $\mathrm{TbTi_3Bi_4}$.
However, we noticed that the broad peak in the $\chi (T)$ for $H \parallel a$-axis was absent in the $\chi(T)$ of $\mathrm{La_{0.972}Tb_{0.028}Ti_3Bi_4}$, indicating that it is related to magnetic ordering.

Strong anisotropic, isothermal magnetization of $\mathrm{La_{0.972}Tb_{0.028}Ti_3Bi_4}$ at 2 K was displayed in Fig.~\ref{fig4} b.
The magnetization for $H \parallel a$ quickly saturated at 20~kOe, with a saturated magnetization $\mu_{sat}$ of $5.27~\mu_\mathrm{B}$ per $\mathrm{Tb^{3+}}$, which is slightly lower than that for Hund's rule ground state of $\mathrm{Tb^{3+}}$.
This discrepancy may be due to the uncertainty of the Tb concentration or the CEF effect.
In comparison, the magnetization for $H \parallel b$-axis did not show any intention of saturation at 70 kOe.
The strongly anisotropic temperature and field dependence of the magnetization of the dilute alloy of $\mathrm{La_{0.972}Tb_{0.028}Ti_3Bi_4}$ shows that the CEF effect plays a critical role in the extreme anisotropy in $\mathrm{TbTi_3Bi_4}$.

\subsection{Magnetic anisotropy in the AFM state}

\begin{figure}[!htbp]
	\centering
	\includegraphics[width=\linewidth]{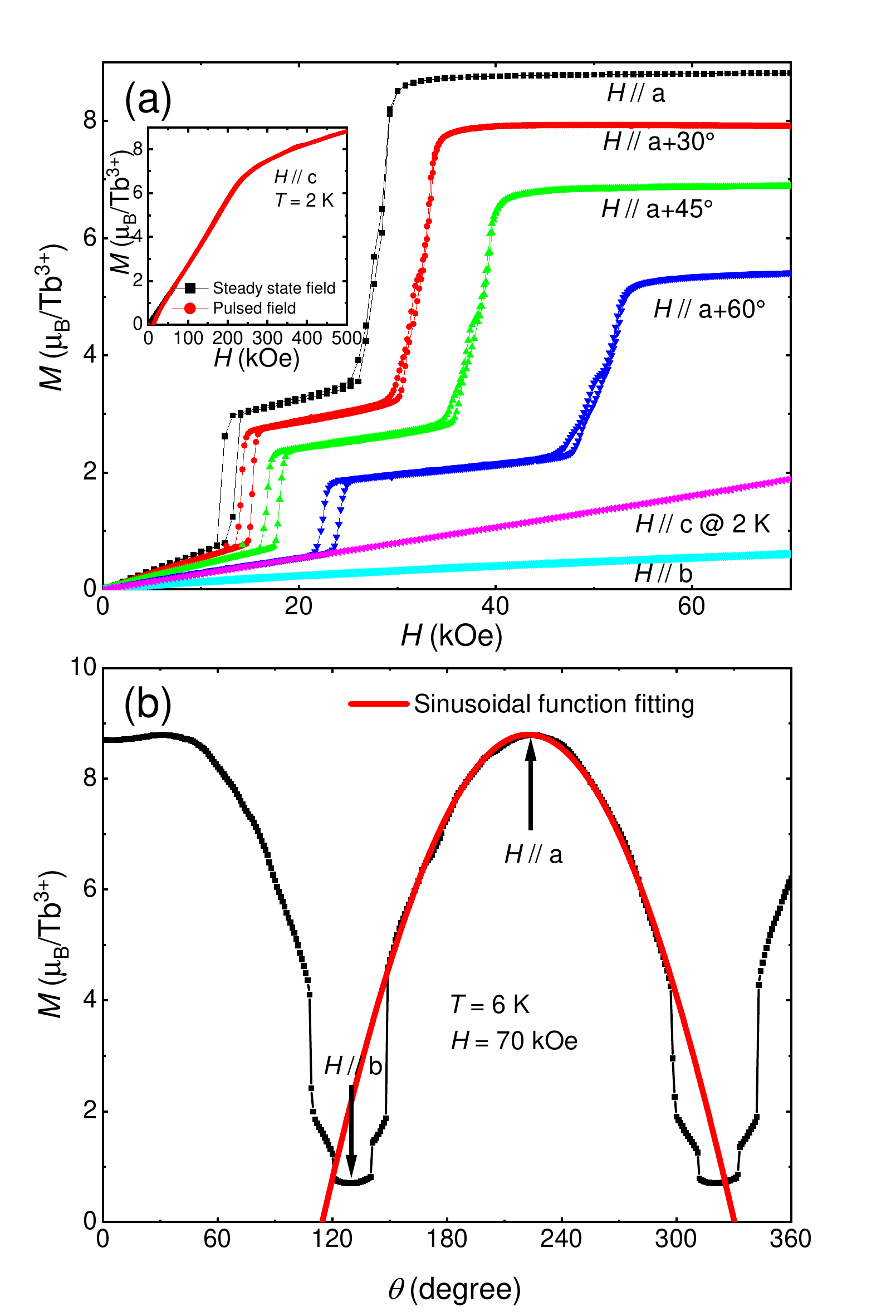}
	\caption{\label{fig5} Magnetic anisotropy of $\mathrm{TbTi_3Bi_4}$ at low temperatures. (a) Field-dependent magnetization when the magnetic field aligns in the $ab$-plane along different directions and parallel to the $c$-axis direction at 2~K, respectively. Inset: Magnetization in a strong, pulsed magnetic field along the $c$-axis. (b) In-plane angle-dependent magnetization at 6~K in a magnetic field being 70~kOe. The red curve represents the result of sinusoidal fitting. }
\end{figure}

The extreme magnetic anisotropy and MMTs of $\mathrm{TbTi_3Bi_4}$ in the AFM state are demonstrated in the low-temperature field-dependent magnetization when the field is applied along the $a$-, $b$-, and $c$-axes, as shown in Fig.~\ref{fig5} a.
The $M(H)$ curves for $H \parallel b$- and $c$-axes exhibit linear behavior and do not saturated up to 70 kOe.
The magnetization along the $c$-axis was also measured in a pulsed magnetic field up to 500~kOe, as seen in the inset of Fig.~\ref{fig5} a, which was calibrated with the low steady-state field data measured in MPMS.
The $M(H)$ curve has a slope change of about 200 kOe but has not saturated with 500 kOe.
On the other hand, the magnetization for $H \parallel a$ shows a step-like feature with the value of about 3 $\mu_{\mathrm{B}}$/$\mathrm{Tb^{3+}}$ when the field is between 14 and 26 kOe, and then saturates at 9 $\mu_{\mathrm{B}}$/$\mathrm{Tb^{3+}}$ when the field is higher than 30 kOe.
Narrow hysteresis loops are observed between the steps, indicating that field-induced MMT is a first-order phase transition.

The magnetization measurements were also performed when the direction of the magnetic field is along a tilted angle deviated from the $a$-axis to the $b$-axis.
As shown in Fig.~\ref{fig5} a, these $M(H)$ isothermals present as a series of step-like curves that gradually evolve from the shape of $M(H)$ for $H \parallel a$ to a linear $M(H)$ for $H \parallel b$.
When the tilted angle $\theta$ changes from $0^{\circ}$ to $60^{\circ}$, the transition and saturated field increase while the magnetization values of the step and the saturation follow the relation of $M\cos(\theta)$ (see in Fig.~\ref{fig5} b).
These features reveal that only the magnetic field component along the $a$ direction plays a role in the MMTs, and the magnetic moments are all along the $a$ direction.

To visualize this extreme one-dimensional anisotropy, we took measurements of the in-plane magnetization at 6 K and 70 kOe, with the angle dependence of $\theta $, as shown in Fig.~\ref{fig5} b where arrows indicate the positions of $H \parallel a$- and $b$-axes.
The change of magnetization with angle reflects a two-folded symmetry, with two step-like features in the $M(\theta)$ curve at $H=70$~kOe, which corresponds to the two MMTs as shown in Fig.~\ref{fig5} a.
When the direction of field tilts away from the $a$-axis, the magnetization initially changes as a sinusoidal function, which is the projection of the saturated magnetic moment along the $a$ direction, represented as a red fitting curve.
When the angle is about 70 degrees away from the $a$-axis, the magnetization suddenly drops to a lower value, followed by a second plunge until it approaches the low value when the direction of the field is along the $b$-axis.
The $M(\theta)$ curves are consistent with the $M(H)$ isothermal and confirm the extreme quasi-one-dimensional magnetic anisotropy in $\mathrm{TbTi_3Bi_4}$.

\begin{figure}[!htbp]
	\centering
	\includegraphics[width=\linewidth]{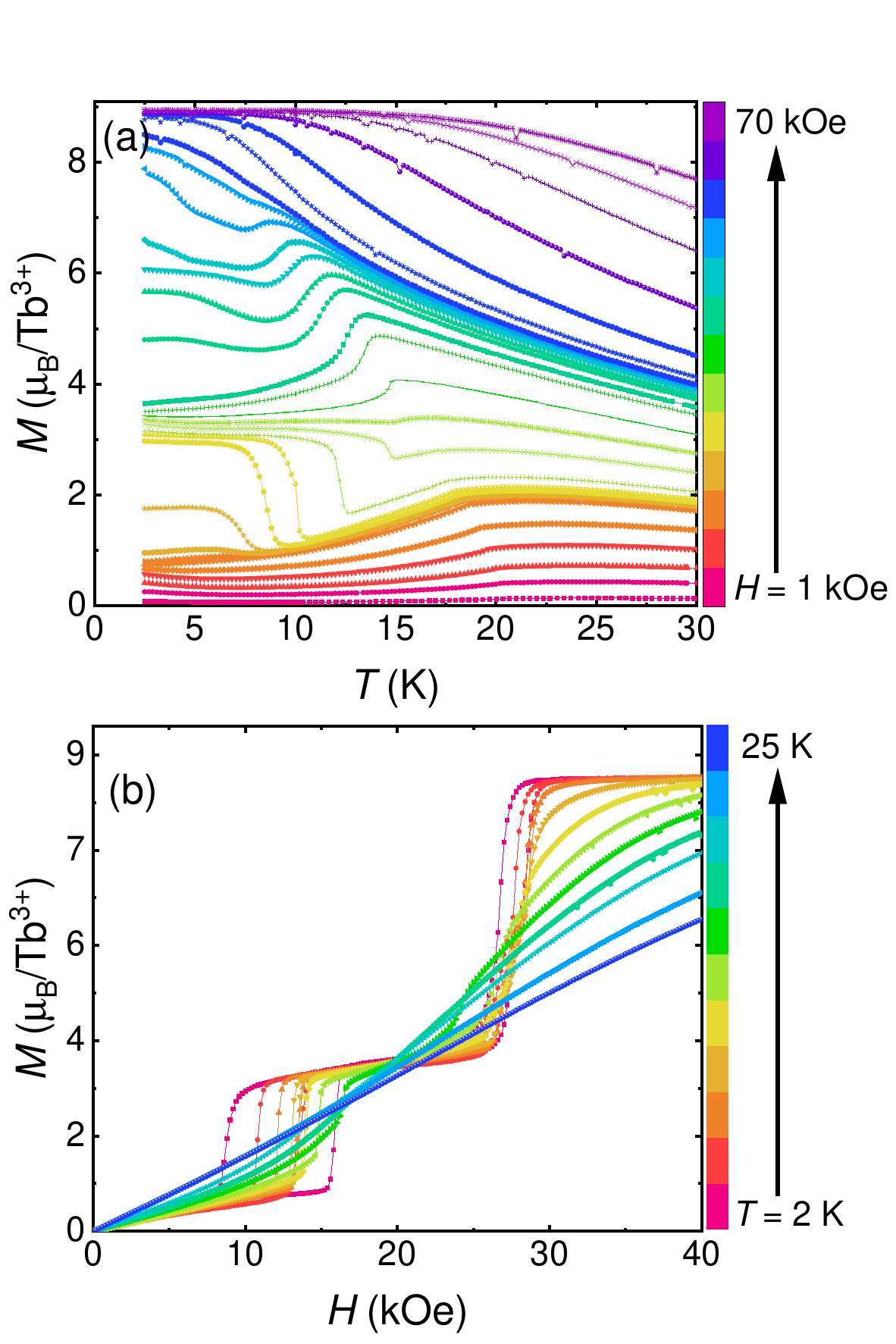}
	\caption{\label{fig6} Magnetization of $\mathrm{TbTi_3Bi_4}$ along the direction of the $a$-axis (a) with respect to the temperatures in different magnetic fields and (b) with respect to the fields at different temperatures. }
\end{figure}

We also performed magnetization measurements in various temperatures and fields along the direction of the $a$-axis to explore the MMTs and intermediate states of $\mathrm{TbTi_3Bi_4}$.
As shown in Fig.~\ref{fig6} a, with the increase of the magnetic field, the broad maximum at 25 K in the $\chi(T)$ curve and the slope change at $T_\mathrm{N}$ being 20.4 K move to low temperature and gradually merge in a field of 22.5 kOe.
The low-temperature magnetization shows a sudden increase and starts to cluster at about $3~ \mu_\mathrm{B}$/f. u. in a field of 15 kOe, corresponding to the first-stage meta-magnetic transition.
The slope change at $T_\mathrm{N}$ is gradually suppressed to lower temperatures with an increase of the field, until it merges with the sudden increase at 22.5 kOe to become a cusp, which indicates that the paramagnetic state has directly changed to the intermediate state.
 When the field is higher than 26 kOe, the low-temperature magnetization increases again and the cusp is suppressed to lower temperature with an increase of the field.
No feature indicates any transition when the field is above 29 kOe and the magnetization starts to saturate.

\begin{figure}[!htbp]
	\centering
	\includegraphics[width=\linewidth]{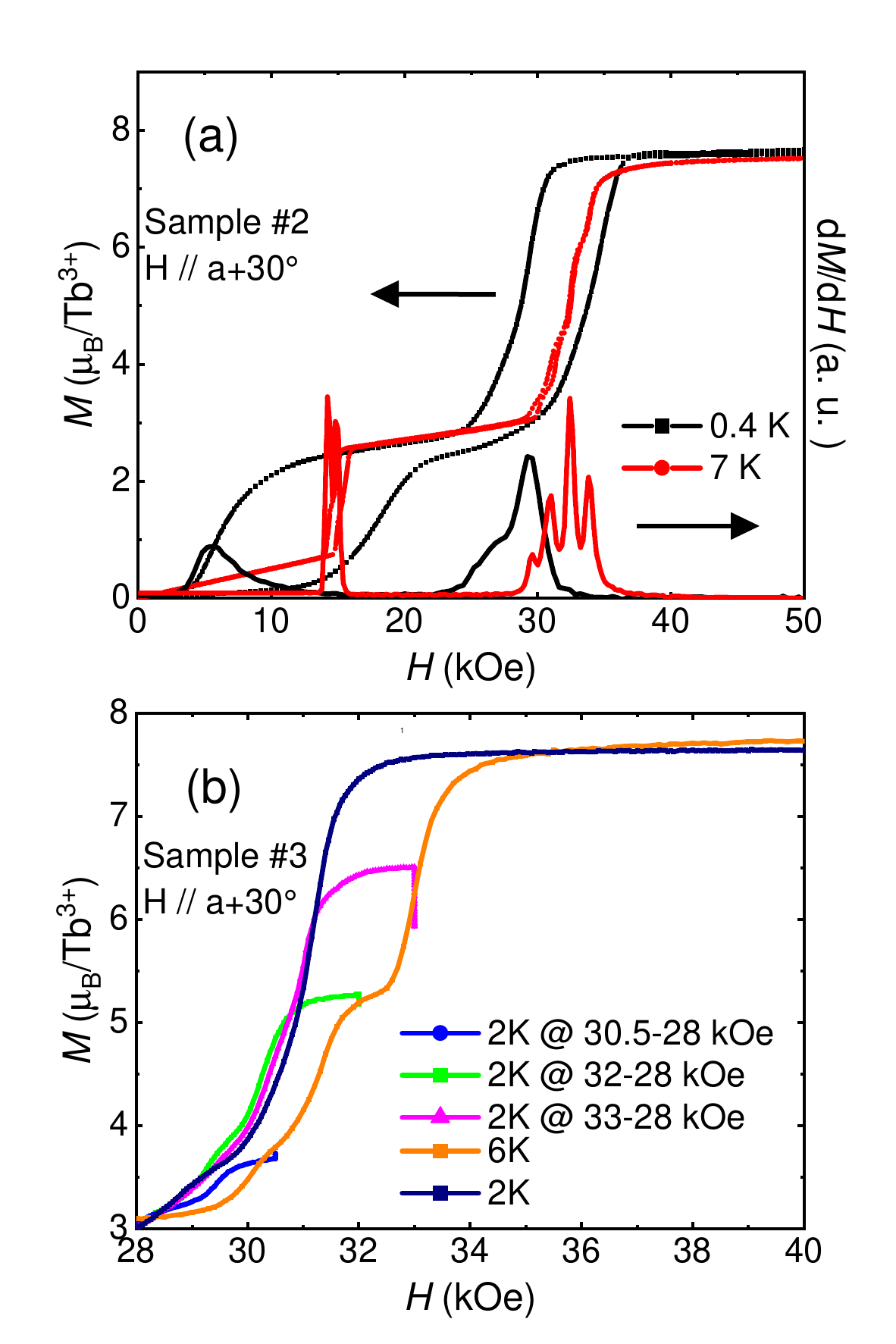}
	\caption{\label{fig7} Existence of other magnetization plateaus in different samples. (a) Left axis: magnetization of $\mathrm{TbTi_3Bi_4}$ for Sample \#2.  Right axis: d$M(H)$/d$H$, the black and red curves indicate temperatures at 0.4~K and 7~K, respectively. (b) Magnetization of $\mathrm{TbTi_3Bi_4}$ for Sample \#3. }
\end{figure}

Figure \ref{fig6} b shows a series of magnetization isothermals from 2 K to 25 K, confirming the evolution of the magnetic states.
Both field-induced MMTs are accompanied by large hysteresis loops at 2 K. As temperature increases, the hysteresis loops become narrower until they change to a spin-flip-like transition at 10 K.

Notably, the series $M(H)$ curves at 6 K in Fig.~\ref{fig5} a have a slight but visual cusp during the process of the second MMT.
The $M$ value of this cusp equals the 2/3 of the saturated magnetization. We realize that these cusps likely represent some meta-stable magnetic structures which can be stabilized at base temperature.

To showing this, we firstly measured the magnetization for Sample \#3 at 6 K when the field swept from 40 kOe to 28 kOe. Here we observed several small plateaus between the 1/3 and saturated magnetization. Then we swept the field from 40 kOe to an intermediate field, and dropped the temperature to 2 K at this field and then swept the field to 28 kOe. These curves are shown in Fig.~\ref{fig7} b with the comparison of the $M(H)$ curve at 2 K. We notice that when the intermediate field is close to the fields of the small plateaus, like the blue and green curves, the magnetization plateau is frozen at 2K. Instead, when the intermediate field is away from the fields od the plateaus, like the scarlet curve, the magnetization drifts away during the cooling. These magnetization processes are not like that for a standard ferromagnet with a hysteresis loop, and indicates that the small plateaus are likely some particular magnetic structures.

\section{Analysis and discussion}

\begin{figure}[!htbp]
	\centering
	\includegraphics[width=\linewidth]{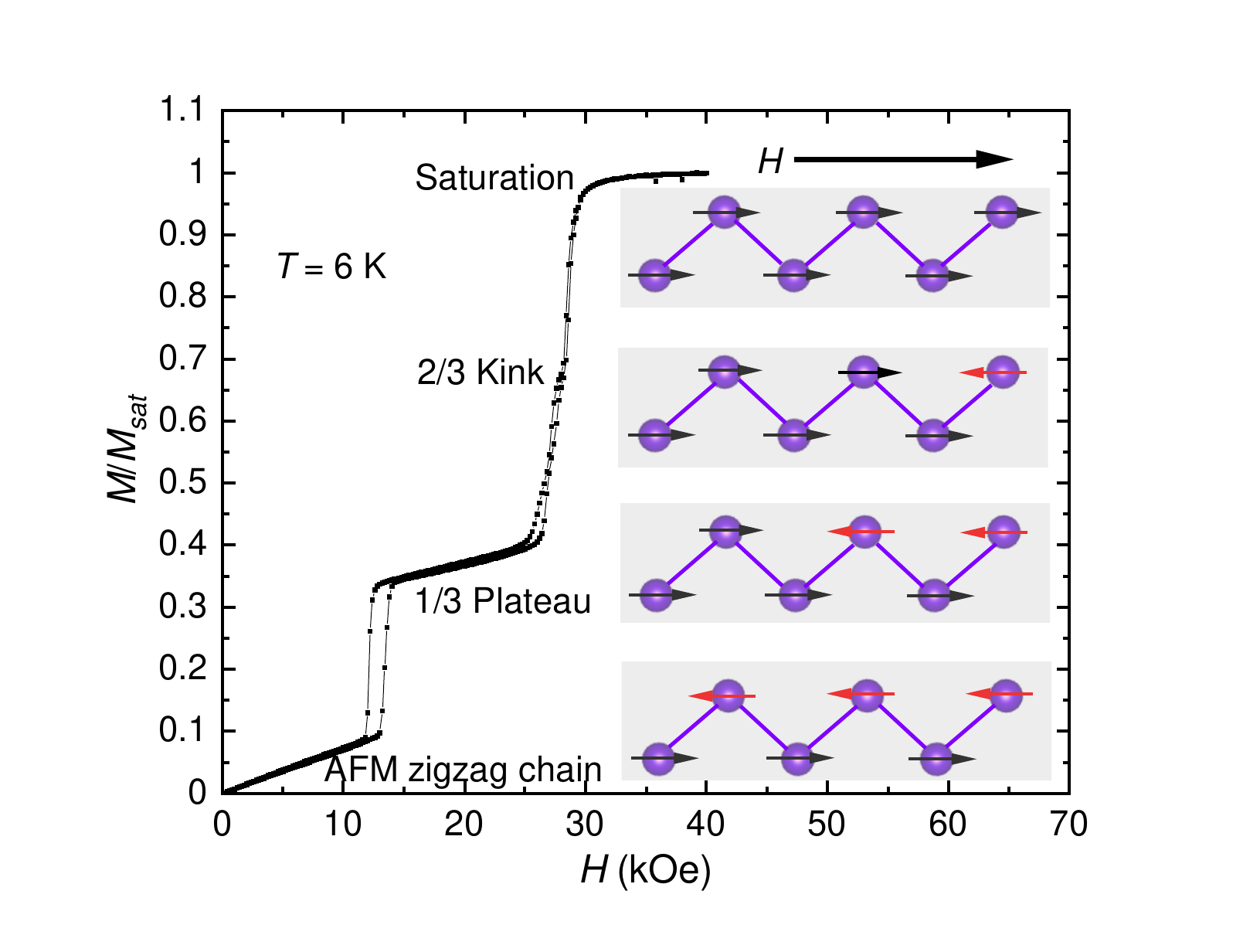}
	\caption{\label{fig8} Schematic of the possible magnetic structure of zigzag chain. The purple atoms are Tb atoms and the arrows indicate the direction of their spin polarization. }
\end{figure}

To better understand how CEF and magnetic interaction impact the propreties of $\mathrm{TbTi_3Bi_4}$, we can compare it to the magnetic properties of isostructural $\mathrm{GdTi_3Bi_4}$ \cite{doi:10.1021/acs.chemmater.3c02289} which lacks CEF because the Hund's rule ground state of Gd$^{3+}$ takes the form of $S_7^{7/2}$.
The previous study showed that $\mathrm{GdTi_3Bi_4}$ presented an AFM ordering at $T_\mathrm{N}$ = 13~K \cite{doi:10.1021/acs.chemmater.3c02289}, much lower than the $T_\mathrm{N}$ of $\mathrm{TbTi_3Bi_4}$ (20.4~K).
The Weiss temperatures ($\theta_{\mathrm{P}}$) of $\mathrm{GdTi_3Bi_4}$ of in-plane and out-of-plane magnetic susceptibilities at high temperatures are positive (10 and 13 K), compared with the negative average $\theta_{\mathrm{P}}$ being -13.6~K for $\mathrm{TbTi_3Bi_4}$.
This difference indicates that the CEF plays a crucial role in $\mathrm{TbTi_3Bi_4}$.

Based on this analysis, we infer a plausible magnetic ordering of $\mathrm{TbTi_3Bi_4}$.
Due to the strong CEF effect, the moment associated with the Tb$^{3+}$ ion presents as an Ising moment, which is constrained in the chain direction ($a$) even at relatively high temperatures.
Because the magnetic interactions in the chain are crucial, the zigzag chain of the Tb moments intends to form a short-range one-dimensional AFM ordering when the temperature decreases.
This short-range ordering occurs at 25~K, corresponding to the broad maximum in the magnetic susceptibility along the direction of the $a$-axis.
The $T_\mathrm{N}$ at 20.4~K represents the transition from short range ordering to three-dimensional ordering. Therefore, the magnetic entropy released in this transition is much less than the ordinary magnetic ordering for the other $R\mathrm{Ti_3Bi_4}$ compounds, which interprets the narrow $\lambda$-shape peak in the heat capacity of $\mathrm{TbTi_3Bi_4}$ (see Fig.~\ref{fig2} c).

The frustrated magnetic interactions in this Ising-like zigzag chain naturally lead to MMTs and fractional magnetization plateaus in an external magnetic field \cite{Okunishi,PhysRevB.68.224422}.
Figure~\ref{fig8} presents the normalized magnetization at 6~K when the field is applied along the direction of the $a$-axis.
The first MMT at 13 kOe reaches a plateaus of $M/M_{sat} = 1/3$, while the second MMT reaches saturation at about 30.6 kOe. The cusp at 28.2 kOe has the $M/M_{sat}$ value close to 2/3.
As shown in the schematic in Fig.~\ref{fig8}, the AFM state of the zigzag chain is composed by the anti-parallel magnetic moments of the nearest neighbor atoms, which are indicated by red and black arrows, respectively.
One-third of the moments in one layer have flipped in the first MMT, leading to a 1/3 plateau state.
This 1/3 state of the zigzag chain has threefold degeneracy in the Ising limit, and a spontaneous symmetry breaking must occur during the transition.
The 2/3 kink likely corresponds to the change of two-third of the moments in one layer flipping.

\begin{figure}[!htbp]
	\centering
	\includegraphics[width=\linewidth]{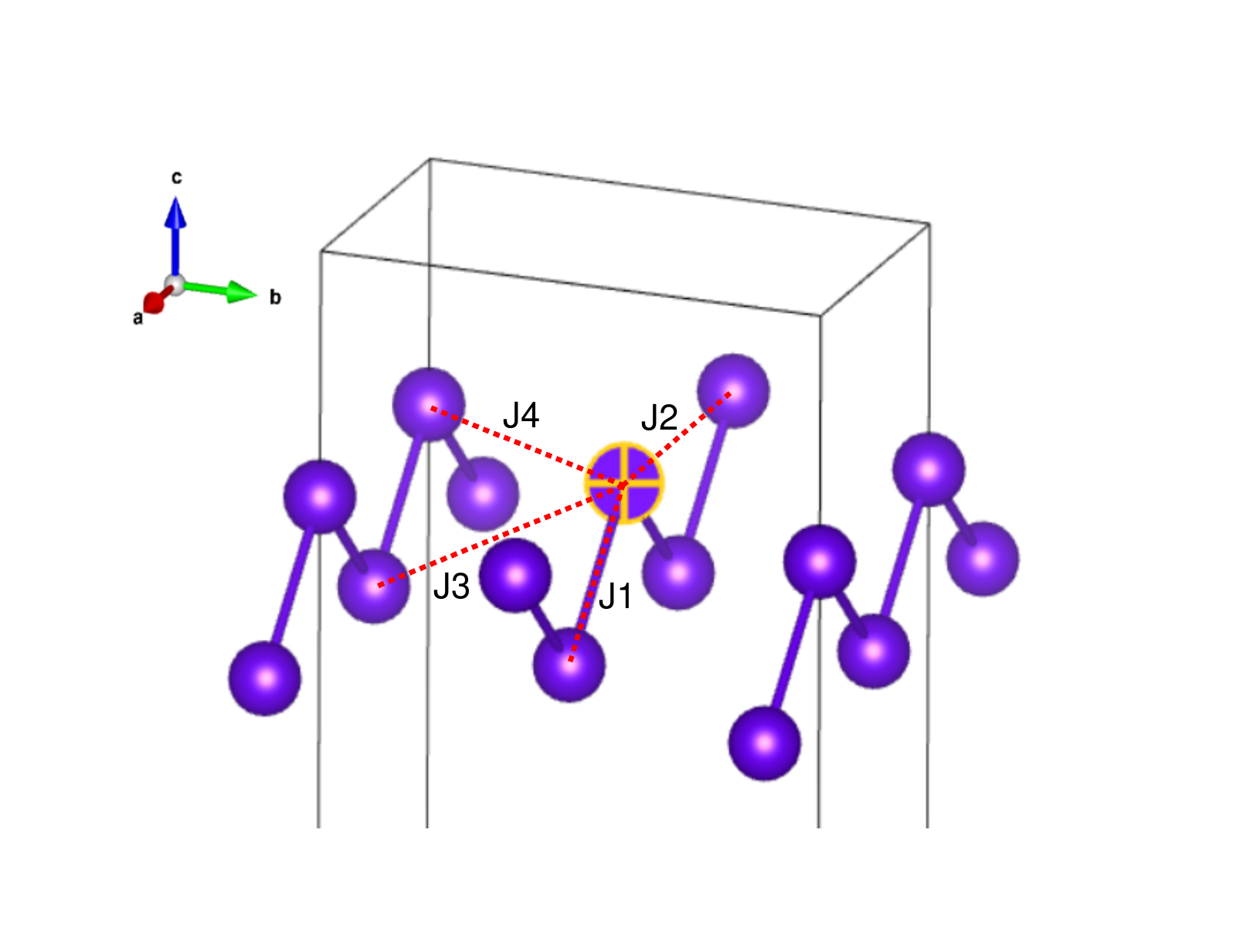}
	\caption{\label{fig9} Schematic of the magnetic interactions between the Tb moments.}
\end{figure}

The AFM zigzag chains can construct different three-dimensional magnetic structures via the interactions parallel in the $b$-axis between the chains.
Because $\mathrm{TbTi_3Bi_4}$ is a metallic system with large local moments, both the short-range dipolar interaction and the long-range Ruderman-Kittel-Kasuya-Yosida (RKKY) interaction play essential roles.
As shown in Fig. \ref{fig9}, each Tb atom has two nearest neighbors (NNs) in the chain with a distance of 3.97 $\mathrm{\AA}$.
Due to the staggered arrangement of the chains along the $b$ direction, the next-nearest neighbors (NNNs) are the two atoms in the next chains in different planes.
There are two next-next-nearest neighbors (NNNNs) in the chain in the plane and four next-next-next-nearest neighbors (NNNNNs) in the next chains in the plane.
Because the NNN and NNNN and NNNNN distances are close (5.83 $\mathrm{\AA}$, 5.89  $\mathrm{\AA}$ and 5.95  $\mathrm{\AA}$), the magnetic interaction $J_1$ in the chain for the NN is anticipated to the strongest. In contrast, the inter-chain interaction $J_3$ is likely comparable with the intra-chain interaction $J_2$ for the NNN and $J_4$ for the NNNNN.
Due to the competition among these interactions, strong magnetic frustration may lead to an FM or stripe order in the plane.
Future spectrum measurements, such as neutron diffraction, are needed to identify the three-dimensional magnetic structure.

A field-induced 1/3 magnetization plateau is commonly observed in frustrated Ising spin systems with a triangular lattice \cite{AHonecker1999,AASLAND1997187,miyashita1986magnetic,PhysRevX.10.011007,GIGNOUX1993139,BALL1992343,PhysRevB.107.094414}.
Depending on the spin frustration of the two-dimensional triangular lattice, the magnetic moments form an up-up-down ($uud$) arrangement in the plane. In contrast, the magnetic moments remain FM in the one-dimensional chain of the vertical triangular lattice.
On the other hand, the spin zigzag chain in $\mathrm{TbTi_3Bi_4}$ leads to a different mechanism of the 1/3 magnetization plateaus.
Previous studies of the one-dimensional $S = 1/2$ zigzag $XXZ$ chain \cite{Okunishi,PhysRevB.68.224422} have predicted the emergence of 1/3 magnetization plateaus.
This theoretical model considers the competition of NNs and NNNs interactions within the chain. 
The 1/3 plateau state is characterized by $uud$ spin structure, which extends to the Ising limit in the magnetic phase diagram.
It is noteworthy that this model not only predicts the emergence of the 1/3 plateau state but also highlights the accompanying spontaneous breaking of translation symmetry, consistent with our schematic in Fig.~\ref{fig8}.
In addition, the $XXZ$ chain model also predicted that the formation of domain walls may lead to multiple cusps in the magnetization near the 1/3 plateau \cite{Okunishi,PhysRevB.68.224422}.
As shown in Fig.~\ref{fig7} a, the cusps near the 1/3 plateau state vanish at sufficiently low temperatures but develop a complicated behavior as the temperature rises.
Our measurements in Fig.~\ref{fig7} b indicate that more complex hidden states may exist around the 1/3 plateau.
The similarity between the theoretically predicted fractional excitation near the 1/3 plateau and the observed magnetization curves is noteworthy.

\begin{figure}[!htbp]
	\centering
	\includegraphics[width=\linewidth]{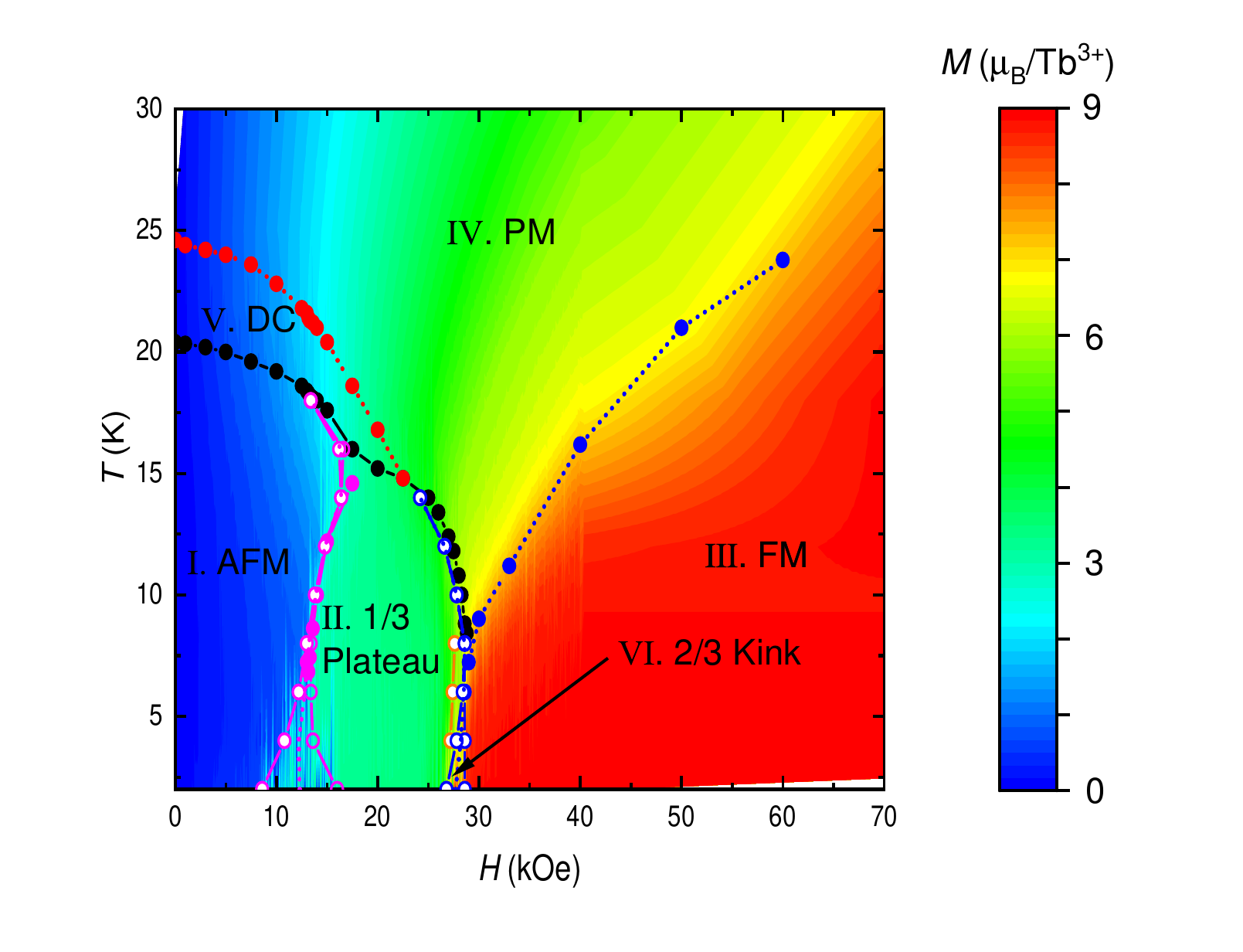}
	\caption{\label{fig10} The magnetic phase diagram of $\mathrm{TbTi_3Bi_4}$. The solid circles are derived from the $M$-$T$ data in Fig.~\ref{fig6}~a, and the hollow circles are derived from the $M$-$H$ data in Fig.~\ref{fig6}~b. }
\end{figure}

The data in Fig.~\ref{fig6} are sorted into the magnetic phase diagram of $\mathrm{TbTi_3Bi_4}$ shown in Fig.~\ref{fig10}. The solid and hollow dots correspond to the data obtained from the $M$-$T$  (Fig.~\ref{fig6} a) and $M$-$H$ (Fig.~\ref{fig6} b) measurements, respectively. The phase-transition temperatures are determined from the peaks in $\mathrm{d}M(T)/\mathrm{d}T$ while the phase-transition fields are determined from the peaks in $\mathrm{d}M(H)/\mathrm{d}H$. The data obtained based on the hysteresis loops are respectively expressed at the same temperature, and their average values are shown as dashed lines in Fig.~\ref{fig10}.
The regions denoting different phases are represented by the Roman numerals from \uppercase\expandafter {\romannumeral1} to \uppercase\expandafter {\romannumeral6}, corresponding to (\uppercase\expandafter {\romannumeral1}) antiferromagnetic state, (\uppercase\expandafter {\romannumeral2}) 1/3 plateau state, (\uppercase\expandafter {\romannumeral3}) ferromagnetic state, (\uppercase\expandafter {\romannumeral4}) paramagnetic state, (\uppercase\expandafter {\romannumeral5}) dimensional crossover and (\uppercase\expandafter {\romannumeral6}) 2/3 kink. It is worth noting that region \uppercase\expandafter {\romannumeral6} is not obvious and is located at the bottom of Fig.~\ref{fig10}, between region \uppercase\expandafter {\romannumeral2} and \uppercase\expandafter {\romannumeral3}.

\section{Conclusions}

To summarize, we have successfully created single crystals of $\mathrm{TbTi_3Bi_4}$, a previously unreported compound.
This material shares the same structure with other $R\mathrm{Ti_3Bi_4}$ members, which is characterized by a bilayer and distorted Ti-based kagome lattices, as well as quasi-one-dimensional Tb-based zigzag chains.
What sets $\mathrm{TbTi_3Bi_4}$ apart is its remarkable AFM ordering at $T_\mathrm{N} = 20.4$ K, significantly higher than other $R\mathrm{Ti_3Bi_4}$ members.
It exhibits properties as a quasi-one-dimensional Ising magnet, with extreme easy-axis magnetic anisotropy due to the CEF effect, aligning the $\mathrm{Tb^{3+}}$ moments along the zigzag chain direction.
Field-dependent measurements reveal a complex magnetic phase diagram, with the $1/3$ plateau state and multiple meta-magnetic transitions between the $1/3$ plateau state and the full-aligned state suggesting the possible formation and motion of magnetic domains, and possible meta-stable states. 
While the magnetism of $\mathrm{TbTi_3Bi_4}$ is partially consistent with previous one-dimensional model, the presence of additional inter-chain interactions requires a more sophisticated model.
Microscopic probes such as neutron and X-ray scattering can provide insight into the magnetic structure and excitations of this fascinating compound.

\section{ACKNOWLEDGEMENTS}
We gratefully thank Gang Chen in ICQM, Jie Ma in Shanghai Jiao Tong University for discussions. We would like to thank Chao Dong, Yinfa Feng, Xinlong Shi in Huazhong University of Science and Technology for their help in our pulsed magnetic field measurements. The work in Peking University was financially supported by the National Key Research and Development Program of China (2021YFA1401900), National Natural Science Foundation of China No.12141002, No.12225401, Innovation Program for Quantum Science and Technology (2021ZD0302600), the Natural Science Foundation of China (Grant No. U22A6005), and the Synergetic Extreme Condition User Facility (SECUF).

\bibliography{TTBcite}

\end{document}